\documentclass[twocolumn,prl,aps]{revtex4}
\usepackage{graphicx}
\usepackage{bm}
\def\Journal#1#2#3#4{{#1} {\bf #2}, #3 (#4)}
\def\NCA{Nuovo Cimento}

\def\NIMA{Nucl. Instr. Meth. A}
\def\NPA{Nucl. Phys. A}
\def\NPB{Nucl. Phys. B}

\def\PLB{Phys. Lett. B}
\def\PRL{Phys. Rev. Lett.}
\def\PR{Phys. Rev.}
\def\PRD{Phys. Rev. D}

\newcommand{\spar}{{\stackrel{\rightarrow}{\Rightarrow}}}
\newcommand{\sant}{{\stackrel{\rightarrow}{\Leftarrow}}}

\begin{document}

\title{Evidence for Quark-Hadron Duality in the Proton Spin Asymmetry $A_1$}

\author{ 
A.~Airapetian,$^{32}$
N.~Akopov,$^{32}$
Z.~Akopov,$^{32}$
M.~Amarian,$^{27,32}$
V.V.~Ammosov,$^{25}$
E.C.~Aschenauer,$^{7}$
H.~Avakian,$^{11}$
R.~Avakian,$^{32}$
A.~Avetissian,$^{32}$
E.~Avetissian,$^{32}$
% ...
\centerline{(The HERMES Collaboration)}
}

\affiliation{
% ...
$^7$DESY Zeuthen, Deutsches Elektronen-Synchrotron, 15738 Zeuthen, Germany \\
% ...
$^{11}$Istituto Nazionale di Fisica Nucleare, Laboratori Nazionali di Frascati, 00044 Frascati, Italy \\
% ...
$^{25}$Institute for High Energy Physics, Protvino, Moscow oblast, 142284 Russia\\
% ...
$^{27}$Istituto Nazionale di Fisica Nucleare, Sezione Roma 1, Gruppo Sanit\`a and Physics Laboratory, Istituto Superiore di Sanit\`a, 00161 Roma, Italy\\
% ...
$^{32}$Yerevan Physics Institute, 375036, Yerevan, Armenia}

\date{
\today}

%
\begin{abstract}
      Spin-dependent lepton-nucleon scattering data have been used to 
      investigate the validity of the concept of quark-hadron duality for 
      the spin asymmetry $A_1$. 
      Longitudinally polarised positrons were scattered off a longitudinally 
      polarised hydrogen target for values of $Q^2$ between 1.2 and 12 
      GeV$^2$ and values of $W^2$ between 1 and 4 GeV$^2$.
      The average double-spin asymmetry in the nucleon resonance
      region is found to agree with that measured in deep-inelastic 
      scattering at the same values of the Bjorken scaling variable $x$.
      This finding implies that the description of $A_1$ in terms of quark
      degrees of freedom is valid also in the nucleon resonance region
      for values of $Q^2$ above 1.6 GeV$^2$.
\end{abstract}

\maketitle

The interaction between baryons and between baryons and leptons can generally 
be described by two complementary approaches: with quark-gluon 
degrees of freedom at high energy, where the quarks are asymptotically free, 
and in terms of hadronic degrees of freedom at low energy, where effects of 
confinement are large.
In some specific cases, where the description in terms of hadrons 
is expected to apply most naturally, the quark-gluon description can also be 
successfully used.  
Such cases are examples of so-called quark-hadron duality.
Bloom and Gilman \cite{BG} first noted this relationship between phenomena 
in the nucleon resonance region and in deep-inelastic scattering (DIS).
Specifically, they observed that the cross section for electro-production of 
nucleon resonances, if averaged over a large enough range of invariant mass $W$
of the initial photon-nucleon system, exhibited the same behavior as 
the cross section observed in the DIS region.
In other words, the scaling limit curve measured as a function of the
variable $x ^{\prime} = x + M^2/Q^2$ in DIS processes at high $Q^2$
and high $\nu$  approximately approaches the smooth curves derived from measurements in the resonance 
region at lower $\nu$ and $Q^2$ (here $x=Q^2/2M\nu$ is the
Bjorken scaling variable, $-Q^2$ is the four-momentum 
transfer squared, $M$ is the proton mass and $\nu$ the energy of the exchanged virtual 
photon in the target rest frame).
\indent
\par
Duality in strong interaction physics was originally
formulated for hadron-hadron scattering \cite{DHSV}: the high-energy 
behavior of amplitudes, described within Regge theory in terms of 
$t$-channel Regge pole exchanges, was related to the behavior of the
amplitudes at low-energy, which are well described by a sum over a few 
$s$-channel resonances \cite{VEN,HAR}. 
In QCD the Bloom-Gilman duality can be interpreted in the language of the
operator product expansion in which moments of structure functions
are expanded in powers of $1/Q$ \cite{DRGP,MEL}. 
The leading terms are associated with non-interacting partons and exhibit 
scaling, while the terms proportional to $1/Q$ involve interactions between 
quarks and gluons. While the first moments of the structure functions depend weakly  on
$Q^2$, this is not true for the higher moments, since 
at large $x$ the scaling violations of structure functions (i.e. the $Q^2$-dependence
for fixed values of $x$) are very large, 
so that a leading order description
in terms of parton distributions is unable to reproduce DIS data.
Therefore, additional terms, which effectively account for higher-order,
higher-twist and target-mass corrections, should be included. It has been
shown that in that way a good description of measured values of $F_2$ structure
function over a wide range of $Q^2$ and $x$ can be obtained \cite{BOYA}.
\indent
\par
Recently, a sample of inclusive unpolarised electron-nucleon scattering 
data on hydrogen and deuterium targets has been analysed to investigate 
the validity of quark-hadron duality \cite{CEBAF}. 
For the proton, it was observed that starting from 
$Q^2 \geq 1.5$ GeV$^2$ duality in the unpolarised structure function 
$F_2$ holds for individual resonance contributions, as well as for the 
entire resonance region 1$\le$$W^2$$\le$4 GeV$^2$.
It is worthwhile to mention that duality in the unpolarised structure function
holds only when comparing the data in the resonance region with phenomenological fits to 
DIS data, while it does not hold when comparing with QCD fits at leading order only.
\indent
\par
In contrast to the extensive study of duality for the unpolarised,
i.e. spin-averaged, photo-absorption cross section, the validity of duality 
has not been investigated for the spin-dependent scattering processes, which are 
related to the spin-dependent photo-absorption cross section. 
Observation of duality for the spin asymmetry $A_1$ is of particular interest 
as it may lead to a complementary means to study the spin structure of the nucleon
at large $x$, which is difficult to measure in the DIS region with
high-statistics.
Since the DIS spin asymmetry $A_1$ has been found to be independent of $Q^2$
for all measured values of $x$, the comparison of this asymmetry in the resonance and in
the DIS regions is straightforward and does not depend on the choice of the
parameterization of DIS cross section data.
\indent
\par
In this Letter the first experimental evidence for quark-hadron duality 
for the proton spin asymmetry $A_1$ is reported.
The data were collected by the HERMES experiment in 1997 with a 27.57 GeV 
longitudinally polarised positron beam incident on a longitudinally 
polarised hydrogen gas target internal to the HERA lepton storage ring at DESY.
The positrons in the HERA ring are transversely polarised by emission of
synchrotron radiation \cite{SOTE}.
Longitudinal polarisation is obtained by using spin rotators located upstream 
and downstream of the HERMES experiment \cite{BA}.
The beam polarisation was measured continuously using Compton backscattering of
circularly polarised laser light \cite{POL}. The average beam polarisation for 
the analysed data was 0.55 with a relative systematic uncertainty of 3.4\%.
\indent 
\par
The HERMES polarised target \cite{TAR} consists of polarised atomic hydrogen 
gas confined in a storage cell, fed by an atomic-beam source of 
nuclear-polarised hydrogen based on Stern-Gerlach separation \cite{SG}. 
The nuclear polarisation of the atoms and the atomic fraction are continuously 
measured with a Breit-Rabi polarimeter \cite{BRP} and a target gas analyser. 
The average target polarisation for the analysed data was 0.88 with a relative 
systematic uncertainty of 4.7\% 
\cite{G1P}.
\indent 
\par
Scattered positrons were detected by the HERMES spectrometer, described in 
Ref.~\cite{SPE}.
For all detected positrons the angular resolution was better than 0.6 mrad,
the momentum resolution was better than 1.6\% aside from bremsstrahlung tails,
and the $Q^2$-resolution was better than 2.2\%.
\indent
\par
In addition to the constraints of the acceptance of the HERMES spectrometer, 
the kinematic requirements for the analysis in the nucleon resonance region 
were: $1 \le W^2 \le 4$ GeV$^2$, and $1.2 \le Q^2 \le 12$ GeV$^2$.
The corresponding $x$ range was 0.34$<$$x$$<$0.98.
After applying data quality criteria, about 120,000 events remained.
\indent 
\par
The evaluation of the longitudinal asymmetry $A_{\|}$ is based 
on the ratio of the luminosity weighted (i.e. normalized) 
count rates using the following formula:
$$
A_{\|}= \frac{N^\sant L^\spar - N^\spar L^\sant}
             {N^\sant L_{\mathrm{P}}^\spar + {N^\spar L_{\mathrm{P}}^\sant}},
$$
where $N$ is the number of detected scattered positrons, $L$  is the 
integrated luminosity corrected for dead time and $L_{\mathrm{P}}$ is the integrated
luminosity corrected for dead time and weighted by the product of the 
beam and target polarisations. 
The superscript $\spar$ $(\sant)$ refers to the orientation of the target spin 
parallel (anti-parallel) to the positron beam polarisation.
\indent
\par
The limited $W$ resolution in the resonance region ($\delta W \approx$ 240 
MeV) does not allow individual nucleon resonances to be distinguished
nor the DIS ($W >$ 2 GeV) and resonance ($W \le$ 2 GeV) regions to be completely separated.
To evaluate the smearing correction and the contaminations in the 
resonance region from the elastic and deep-inelastic regions, these effects 
were studied using a simulation of events from elastic, resonance, 
and deep-inelastic processes. 
The parameterisations of these contributions were taken from 
Refs.~\cite{BIL,BOD,nmcp8}.
The contamination from elastic and DIS events in the resonance region varies
from 9.7\% to 3.3\% and from 9.5\% to 18.7\%, respectively, with $Q^2$ ranging 
from 1.2 to 12 GeV$^2$.
\indent
\par
The virtual photo-absorption asymmetry $A_1$ is proportional to the cross 
section difference $(\sigma_{1/2} - \sigma_{3/2})$, where $\sigma_{1/2}$
and $\sigma_{3/2}$ are the photo-absorption cross sections for total 
helicities 1/2 and 3/2, respectively.
The asymmetry $A_1$ was extracted \cite{GDH} from the measured longitudinal 
asymmetry $A_{\parallel}$ using the relation 
$A_1 = A_{\parallel}/D - \eta A_2$,
where $D$ is the virtual photon depolarisation factor and $\eta$ is a
 kinematic factor \cite{GDHdis}.
It is noted that the quantity $D$ depends on the ratio 
$R=\sigma_L/\sigma_T$ of absorption cross sections for longitudinal and 
transverse virtual photons \cite{WIT}.
The asymmetry $A_2$ is related to the structure function $g_2(x)$
by $A_2$=$\gamma (g_1(x)+g_2(x))/F_1(x)$, where $\gamma^2 = Q^2/\nu^2$.
The asymmetry $A_1$ was calculated under the assumption that 
$A_2$=0.06$\pm$0.16 as obtained from SLAC measurements \cite{E143} at 
$Q^2$=3 GeV$^2$.
\indent
\par
The spin asymmetry in the nucleon resonance region 
$A_1^{\mathrm res}$ is given
in Table I and is shown as a function of $x$ in Fig.~1.
For each value of $x$ the quantity $A_1^{\mathrm res}$ has been averaged 
over $Q^2$. 
The average $Q^2$ ranges from 1.6 GeV$^2$ in the lowest $x$ bin to 
2.9 GeV$^2$ in the highest. 
The total systematic uncertainty of the data is about 16\%, with the
dominant contribution originating from $A_2$ amounting to 14\%. 
This contribution was evaluated using the measured uncertainty of $A_2$ 
quoted above.
The uncertainty of 14\% is also consistent with the assumption that 
$A_2$=0, and the assumption that $A_2$=0.53 $Mx$ $/\sqrt{Q^2}$, which 
describes its behavior in the deep inelastic region \cite{G1P}.
The experimental systematic uncertainty receives a total contribution of about 8\% 
from the following sources.
The resolution smearing effects give contributions up to 5.6\%. They were 
evaluated by comparing simulated results from two very different assumptions 
for $A_1$, a power law ($A_1^{\mathrm res}$=$x^{0.7}$), and a step 
function ($A_1^{\mathrm res}$=-0.5 for $W^2<$1.8 
GeV$^2$ and 
$A_1^{\mathrm res}$=1.0 for $1.8 \leq W^2 \leq 4.0$ GeV$^2$), which is 
suggested by the hypothesis of the possible dominance 
of the $P_{33}$ resonance at low $W^2$ and the $S_{11}$ at higher $W^2$. 
The modification of the depolarisation factor $D$ due to smearing effects was 
also taken into account.
Other contributions are the uncertainties from beam and target polarisation 
(5.3\%) and from the spectrometer geometry (2.5\%).
Contributions from radiative corrections, calculated using the POLRAD code 
\cite{RAD}, gave a contribution of up to 3\% to the systematic uncertainty.
\indent
\par
Also shown in Fig.~1 is the asymmetry $A_1^{\mathrm DIS}$ as measured in DIS 
\cite{G1P,E143,SMC,E155}.
The data in the resonance region are in agreement with those measured in 
DIS.
The data indicate that $A_1^{\mathrm res}$ may exceed the exact spin-flavor SU(6) symmetry 
expectation of 5/9 at $x$=1, being in better agreement with the original 
and long standing prediction of 1 at $x$=1 \cite{CL1}.
This latter prediction is also favored by the measured $A_1^{\mathrm DIS}$ at large $x$ and by 
more recent expectations \cite{IS,BRO,MA}.
The curve in Fig.~1 is a power law fit to the world DIS data at $x>$0.3:
$A_1^{\mathrm DIS}$ = $x^{0.7}$.
This parameterisation of $A_1$ is constrained to 1 at $x$=1 and does not 
depend on $Q^2$, as indicated by experimental data in this range \cite{E155}.
The average ratio of the measured $A_1^{\mathrm res}$ to the DIS fit 
is 1.11 $\pm$ 0.16 (stat.) $\pm$ 0.18 (syst.).

\begin{figure}[h!]
   \begin{center}
   \includegraphics[width=8cm]{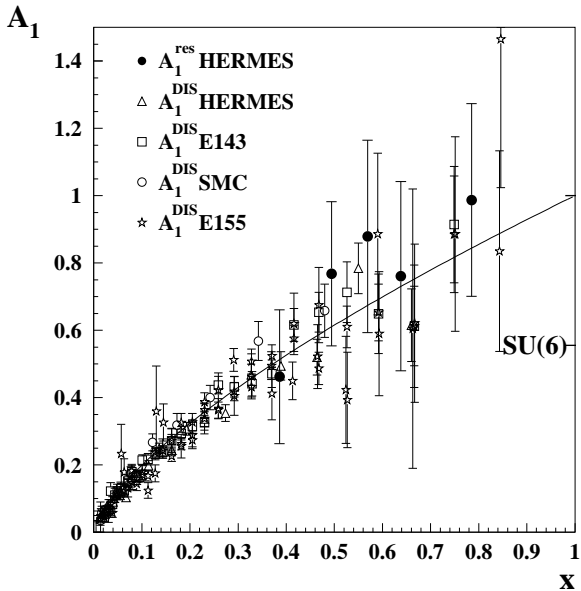}
   \end{center}
\caption{Spin asymmetry $A_1$ as a function of $x$ measured in the resonance
region (full circles).
Error bars represent the statistical uncertainties; the systematic 
uncertainty for the data in the resonance region is about 16\%.
Open symbols are previous results obtained in the DIS region.
The curve represents a power law fit to DIS data at $x>$0.3.} 
\label{Fig. 1}
\end{figure}

Originally, duality was introduced by Bloom and Gilman \cite{BG}
by considering the variable $x^{\prime} = x + M^2 / Q^2$ instead of the 
Bjorken variable $x$, while more recently the Nachtmann variable 
$\xi = 2x/(1+\sqrt{1+\gamma^2})$ ~\cite{NAC} was generally used
for duality studies ~\cite{CEBAF}. 
The latter variable accounts for the effects of the
mass of the target which are not negligible in the nucleon resonance
excitation kinematics. In Table I, the relevant values for the Nachtmann variable $\xi$
are reported together with the ones for the Bjorken variable $x$. 
The difference between the two variables amounts to 
about $10\%$ in the HERMES kinematics and this difference results in a small
target-mass correction of about $5\%$ to the ratio of $A_1^{\mathrm res}$ to  
$A_1^{\mathrm DIS}$.
\indent
\par
These results suggest that the description of the spin asymmetry in terms of quark 
degrees of freedom is valid also in the nucleon resonance region for the $Q^2$-range 
explored by the present experiment.
The evidence for duality in both the spin-averaged and the spin-dependent scattering
processes means that the photo-absorption cross sections for the two 
helicity states ($\sigma_{1/2}$ and $\sigma_{3/2}$) exhibit duality 
separately. 
\indent
\par
It is worth mentioning that the measured spin asymmetry in the resonance
region for $Q^2>$1.6 GeV$^2$, where the asymmetry is dominated by
the $\sigma_{1/2}$ component, is positive and 
has the opposite sign with respect to the one measured in the real photon
limit ($Q^2 =0$), where the helicity asymmetry of leading resonances is dominated by
the $\sigma_{3/2}$ component \cite{MAMI}. Since the measured spin asymmetry in the DIS
region is always positive for any $Q^2$, duality in the spin asymmetry
must break down as $Q^2$ goes to zero. In particular it has been argued 
that duality must fail near $Q^2 \sim$ 0.5  GeV$^2$, where the electric
and magnetic multipoles in the virtual photoabsorption are expected to have 
comparable strengths \cite{CL2}.

\begin{table}[h]
\caption{Spin asymmetry in the nucleon resonance region 
$A_1^{\mathrm res}$ as a function of the Bjorken variable $x$
and of the Nachtmann variable $\xi$. For each value, the average
$Q^2$ is also given. $\delta A_1^{\mathrm res}$
represent the statistical uncertainties; the systematic 
uncertainty for the data is about 16\%.}
\begin{center}
\medskip
\begin{ruledtabular}
\begin{tabular}{cccc}
$x$ & $\xi$ & $\langle Q^2$ $\rangle$
(GeV$^2$) & $A_1^{\mathrm res} \pm \delta A_1^{\mathrm res}$ \\ \hline
0.38 & 0.36 & 1.6 & 0.46 $\pm$ 0.20 \\ 
0.50 & 0.45 & 2.0 & 0.77 $\pm$ 0.21 \\ 
0.57 & 0.51 & 2.3 & 0.88 $\pm$ 0.29 \\ 
0.64 & 0.57 & 2.6 & 0.76 $\pm$ 0.28 \\ 
0.78 & 0.68 & 2.9 & 0.99 $\pm$ 0.29  
\end{tabular}
\end{ruledtabular}
\end{center}
\end{table}

In summary, the first experimental evidence of quark-hadron duality for the
spin asymmetry $A_1 (x)$ of the proton has been 
observed for $Q^2$ between 1.6 GeV$^2$ and 2.9 GeV$^2$. 
The spin asymmetries measured in the nucleon
resonance region at $W^2 \le$ 4 GeV$^2$ have been found to be in agreement with the spin
asymmetries measured in the DIS region at larger $W^2$. 
Target-mass effects are found to be small in the HERMES kinematics.
This experimental finding indicates that the description of the spin
asymmetry in terms of quark degrees of freedom is on average valid also in the
nucleon resonance region within the $Q^2$-range probed by the present 
experiment.
\indent
\par
We gratefully acknowledge the DESY management for its support, the
staffs at DESY and the collaborating institutions for their
significant effort, and our funding agencies for financial support.
This work was supported by the FWO-Flanders, Belgium; the Natural Sciences 
and Engineering Research Council of Canada; the INTAS and ESOP network
contributions from the European Community; the German Bundesministerium
f\"ur Bildung und Forschung; the Deutsche Forschungsgemeinschaft (DFG);
the Deutscher Akademisches Austauschdienst (DAAD); the Italian Istituto  
Nazionale di Fisica Nucleare (INFN); Monbusho International Scientific
Research Program, JSPS, and Toray Science Foundation of Japan; the Dutch
Foundation for Fundamenteel Onderzoek der Materie (FOM); the U. K. Particle 
Physics and Astronomy Research Council; and the U. S. Department of Energy 
and the National Science Foundation.

{}

\end{document}